\documentclass[pdftex,12pt,a4paper]{article}
\usepackage[mathscr]{euscript}
\usepackage{color}
\usepackage[pdftex]{graphicx}
\usepackage{amsmath}
\title{Numerical simulations of stick percolation:
Application to the study of structured magnetorheologial elastomers}
\author{
J.\ L.\ Mietta\textsuperscript{1}, 
R.\ M.\ Negri\textsuperscript{1}, and 
P.\ I.\ Tamborenea\textsuperscript{2}\\
\small{$^1$ INQUIMAE and $^2$ Departamento de F\'isica and IFIBA},\\
\small{Facultad de Ciencias Exactas y Naturales,
       Universidad de Buenos Aires}\\
\small{Ciudad Universitaria, Buenos Aires, ARGENTINA}
}

\begin{document}

\maketitle

\begin{abstract}
In this article we explore how structural parameters of composites
filled with one-dimensional, electrically conducting 
elements (such as sticks, needles, chains, or rods) 
affect the percolation properties of the system.
To this end, we perform Monte Carlo simulations of asymmetric 
two-dimensional stick systems with anisotropic alignments.
We compute the percolation probability functions in 
the direction of preferential orientation of the percolating objects 
and in the orthogonal direction, as functions of the experimental 
structural parameters.
Among these, we considered the average length of the sticks, 
the standard deviation of the length distribution, and the standard 
deviation of the angular distribution. 
We developed a computer algorithm capable of reproducing and verifying 
known theoretical results for isotropic networks and which allows us to go beyond 
and study anisotropic systems of experimental interest. 
Our research shows that the total electrical anisotropy, considered
as a direct consequence of the percolation anisotropy, depends 
mainly on the standard deviation of the angular distribution 
and on the average length of the sticks.
A conclusion of practical interest is that we find that there 
is a wide and well-defined range of values for the mentioned 
parameters for which it is possible to obtain reliable anisotropic 
percolation under relatively accessible experimental conditions 
when considering composites formed by dispersions of sticks, 
oriented in elastomeric matrices.
\end{abstract}

%%%%%%%%%%%%%%%%%%%%%%%%%%%%%%%%%%%%%%%%%%%%%%%%%%%%%%%%%%%%%%%%%%%%%%%%%%%%%%%

\section{Introduction and Experimental Motivation}

Recently, a vigorous interest has arisen in percolating networks built 
out of nano- and micro-dimensional objects (percolating objects), 
such as nanotubes and nanowires for various applications such as 
thin film transistors \cite{gup-beh-lia,zen-xu-she},
flexible microelectronics \cite{koc-mei-gau},
microelectromechanical systems (MEMS) \cite{mie-jor-per,sno-cam-anc}, 
chemical sensors \cite{coll-bra-ish}, and
construction of transparent electrodes for optoelectronic and photovoltaic 
devices \cite{hic-beh-ura,pas-una-kan}.
In particular, if these objects are used as fillers dispersed 
in an elastomeric polymer and then oriented inside the organic 
matrix, then there is an anisotropic internal structure to 
the material.
This anisotropic structure can be obtained in practice following
different methods, like, for example, orienting the filler particles
by means of an external field (electric or magnetic), 
mechanically squeezing a composite material, etc.
A particular case is given by magnetorheological materials,
whose mechanical properties can be modified by externally applied 
magnetic fields. 
An important example of magnetorheological materials are composites 
formed by dispersing magnetic filler particles into an elastomer polymeric 
matrix and then orienting the particles. 
These materials are referred to as magnetorheological elastomers (MRE). 
If, additionally, the filler particles are electrically conducting,
the MRE may also be a conductor depending on the 
properties of the filler and matrix materials, and on the conditions
of synthesis.
A simple procedure to obtain MRE (in film or bulk form) consists of curing the 
filler-elastomer composite in the presence of a uniform magnetic field, 
which induces agglomerations of the filler particles into chain-like structures 
(needles) aligned in the direction of the magnetic field
\cite{mie-rui-ant,ant-jor-per,kch-bos,mie-jor-per}.

A desirable property in an electrically conducting MRE is its electrical anisotropy, 
i.e.\ its ability to conduct an electrical current preferentially 
in a special direction. 
Although we will not tackle here the full problem of the 
relation between percolation and electrical conductivity, 
for our current purposes it will suffice to use the fact that 
there will be a total electrical anisotropy (TEA) in the MRE, 
i.e.\ conduction in only a chosen direction, if there are percolating 
paths formed only in that chosen direction. 
This direction is given by the preferred orientation of the needles, which 
coincides with the direction of magnetic field applied during the curing 
of the material. 
For example, TEA is a crucial property in devices like extended pressure 
mapping sensors \cite{mie-jor-per} and 
Zebra\textsuperscript{\textregistered}-like connectors 
for parallel flip-chip connections \cite{mie-rui-ant}.
If the experimental variables are not properly set during the fabrication of 
the MRE, a material devoid of anisotropy (or with very low anisotropy) 
might be obtained \cite{mie-rui-ant,mie-jor-per,boc-awi}, which would be
unsuitable for these types of applications.

The microscopic structure of an MRE film filled with randomly distributed
magnetic sticks can be characterized by three parameters: 
the average length of the sticks, $\langle\ell\rangle$,
the standard deviation of the length distribution, $\sigma_{\ell}$, and
the standard deviation of the angular distribution, $\sigma_{\theta}$
(around a chosen direction).
As will be seen later, in the experimental samples these 
parameters correspond to a normal 
distribution for the angle $\theta$
and a log-normal distribution for the length $\ell$.
These parameters can be set experimentally by changing the intensity of the 
magnetic field during the curing, the exposure time to the magnetic field 
before starting the thermal curing, the viscosity of the matrix, the amount of 
filler, the magnetic properties of the filler, etc. 
Two of us have performed several experimental studies of these
systems \cite{mie-jor-per,mie-rui-ant,ant-jor-per}, and now
we wish to understand how their structural 
parameters affect the probability of obtaining systems with TEA.
Nevertheless, it can be expected that the results of our modeling 
will also apply to other composite materials with similar internal 
structure.

In magnetorheological elastomers, micro- or nanoscopic objects 
(called alternatively chains, needles, or sticks) are formed by 
magneto-piezo-electric manipulation under appropriate experimental conditions 
at various stages of the synthesis process \cite{mie-rui-ant,mie-jor-per,boc-awi}.
These sticks formed by the filler material can be considered as 
quasi-one-dimensional objects, so that the nanostructured MRE 
as a whole can be analyzed in terms of networks of percolating sticks.  
Percolation can be studied either in bulk (3D) or planar (2D) geometry.
For our purposes (i.e.\ studying the TEA) both situations provide 
useful models, and we will therefore consider the latter, which allows 
us to explore numerically larger systems.
In rectangular or square MRE films, by definition there 
is a spanning cluster if there is at least 
one connecting path between two opposing electrical contacts 
(located on opposite edges of the film) formed by intersecting sticks.
In order to relate electrical conduction to percolation concepts,
here we will adopt the criterion that there is TEA if and only if 
there is a spanning cluster in one direction and not in the other.
Thus, we adopt here the well-known two-dimensional model of percolating 
sticks, which is an example of a continuous percolation model 
\cite{yoo-cho-kim,li-zha,kim-yun-yoo,zez-sta-bel,sta-aha,mat-lin-ram,
amb-gri-bal,nig-gri-rys}
in order to study the dependence of the TEA on the structural parameters 
$\sigma_{\theta}$, $\langle\ell\rangle$, and $\sigma_{\ell}$.
We follow the approach of studying the percolation probability functions 
in the direction of application of the curing magnetic field and 
in the orthogonal direction, as functions of the experimental parameters 
$\sigma_{\theta}$, $\sigma_{\ell}$, and $\langle\ell\rangle$. 
For this, we have developed a computer algorithm that is able to reproduce 
and verify known theoretical results for isotropic networks and which allows 
us to go beyond and study anisotropic systems of experimental interest 
(for which very limited results are available in the literature.
Most available studies deal with square systems, with uniform stick 
length and uniformly distributed angular anisotropy 
\cite{zen-xu-she,yoo-cho-kim,lin-lee-an}).

%%%%%%%%%%%%%%%%%%%%%%%%%%%%%%%%%%%%%%%%%%%%%%%%%%%%%%%%%%%%%%%%%%%%%%%%%%%%%%%

%\section[simulations]{Stick-Percolation Simulations}

%%%%%%%%%%%%%%%%%%%%%%%%%%%%%%%%%%%%%%%%%%%%%%%%%%%%%%%%%%%%%%%%%%%%%%%%%%%%%%%
\section{Algorithm of Stick-Percolation Simulations}

We study two-dimensional stick percolation by means of computer simulations. 
The percolation probabilities are obtained by repeating a large number of times
a percolation experiment (a ``realization'').
Each experiment consists of, starting with an empty rectangular 
(sides $L_x, L_y$, aspect ratio $r=L_x/L_y$) or square (side $L=L_x=L_y$)
box, adding a desired number 
$\mathscr{N}$ of sticks of length $\ell$ (either a fixed length 
or statistically distributed values),
and determining whether a spanning cluster exists between the vertical
and/or horizontal faces of the box.
The orientation of a stick is given by an angle $\theta$ with respect to a 
horizontal axis, also generated randomly, with in general 
$-\pi \le \theta \le \pi$.
Depending on the characteristics of the physical system that one wishes to
simulate, different statistical distributions for the lengths $\ell$
and the angles $\theta$ can be adopted.
For example, the simplest and most usually studied model is the one
with a uniform length for all sticks and isotropic angular distribution.
Once the sticks of a given realization (our ``percolating objects'') 
have been generated, we need to first determine which sticks intersect.
Let $A_i$ and $B_i$, for $i=1,\ldots,\mathscr{N}$, denote the endpoints of 
the sticks.
For algorithmic purposes, the sticks can be seen as vectors, 
$\overrightarrow{A_iB_i}$.
Consider a subsystem consisting of only two sticks, $\overrightarrow{A_1B_1}$
and $\overrightarrow{A_2B_2}$.
It can be demonstrated that they intersect if the following conditions 
are simultaneously satisfied \cite{ber-che-van}
\begin{eqnarray}
 (\overrightarrow{A_1B_1} \times \overrightarrow{A_1A_2}) \cdot
 (\overrightarrow{A_1B_1} \times \overrightarrow{A_1B_2}) \leq 0, \nonumber \\
 (\overrightarrow{A_2B_2} \times \overrightarrow{A_2B_1}) \cdot
 (\overrightarrow{A_2B_2} \times \overrightarrow{A_2A_1}) \leq 0.
\end{eqnarray}
The intersection pattern of the $\mathscr{N}$ sticks is explored pairwise 
with these conditions, and an $\mathscr{N} \times \mathscr{N}$ intersection 
matrix $J$ is formed such that $J_{ij}=1$ if the sticks $i$ and $j$ intersect, 
and $J_{ij}=0$ if they do not.
The matrix $J$ is then used as the matrix associated to an intersection graph, 
and a Deep First Search algorithm \cite{erc} is implemented to evaluate 
the presence of spanning clusters connecting the opposite 
edges of the system. 
If out of $n$ realizations of the system $k$ of them possess at least one
spanning cluster, then $k/n$ is an unbiased estimator of the percolation
probability (probability that a system has at least one cluster, referred 
here as $\wp$), i.e. $k/n\rightarrow\wp$ with $n\rightarrow\infty$.
$X\equiv\ k/n$ is a binomially distributed random variable with mean $E[X]=\wp$ 
an variance $\text{Var}[E]=[\wp \times (1-\wp)]/n$ \cite{dev}.
Then, for acceptable accuracy (estimator error lower than 2\%), at least 
$10^3$ Monte Carlo realizations were performed, yielding comparable or 
even better statistics than those found in recent 
studies \cite{zen-xu-she,lin-lee-an}.
We implemented our algorithm in a computer program written in SAGE.

\begin{figure} %1
\centerline{\includegraphics[width=12cm]{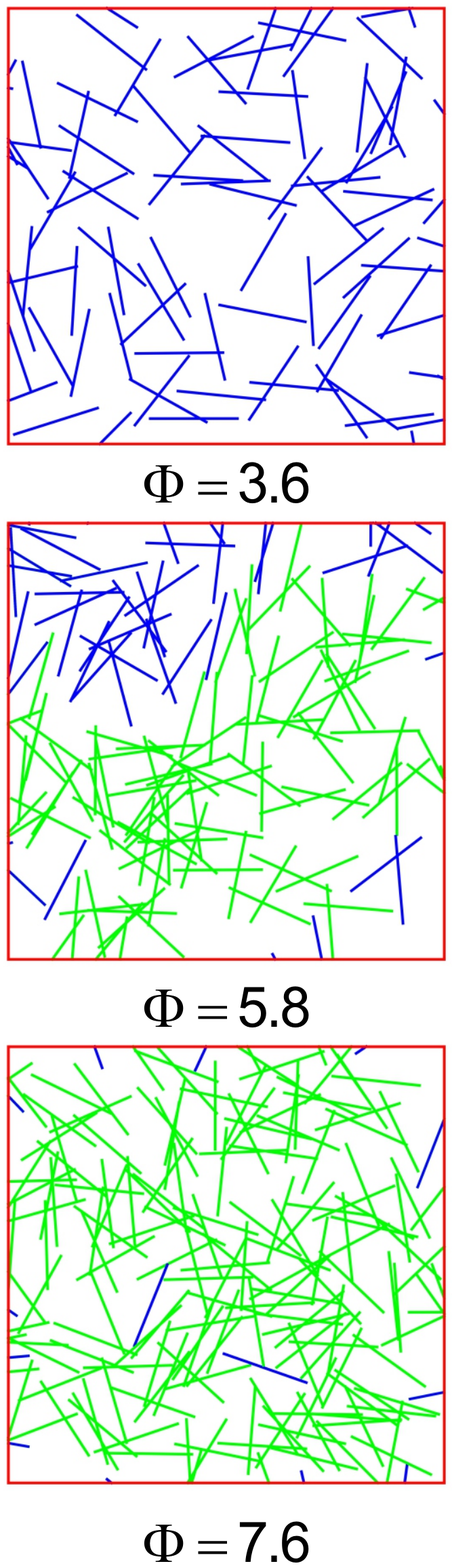}}
\caption{Examples of random stick systems in a square two-dimensional box
of side $L=5$ with isotropic angular distribution and stick length 
$\ell=1$ (all lengths are expressed in the same arbitrary units), 
for different stick densities. 
%$\Phi=3.6\, \text{chains/mm}^2$, well below the critical density,
%$\Phi=5.8\, \text{chains/mm}^2$, slightly above the critical density, and 
%$\Phi=7.6\, \text{chains/mm}^2$.
The green (color online) sticks belong to a horizontal spanning cluster and the 
blue ones do not.}
\label{fig:square_images}
\end{figure}

Figure \ref{fig:square_images} shows three examples of random stick systems 
in a square two-dimensional box of side $L=5$
with isotropic angular distribution and fixed stick length $\ell=1$ 
(lengths expressed in the same arbitrary units), 
for different stick densities, $\Phi=\mathscr{N}/L^2$, especially 
chosen in order to display the percolation behavior, well below, 
slightly above, and well above the percolation critical density.
For the low density there are no spanning clusters.
For the described system, for the intermediate density there is one 
spanning cluster which contains about 60\% of the sticks, and for the high
density most of the sticks participate in the spanning cluster.

%%%%%%%%%%%%%%%%%%%%%%%%%%%%%%%%%%%%%%%%%%%%%%%%%%%%%%%%%%%%%%%%%%%%%%%%%%%%%%%

\section{Finite-Size Scaling for Square Isotropic Systems}

\begin{figure} %2
\centerline{\includegraphics[width=8cm]{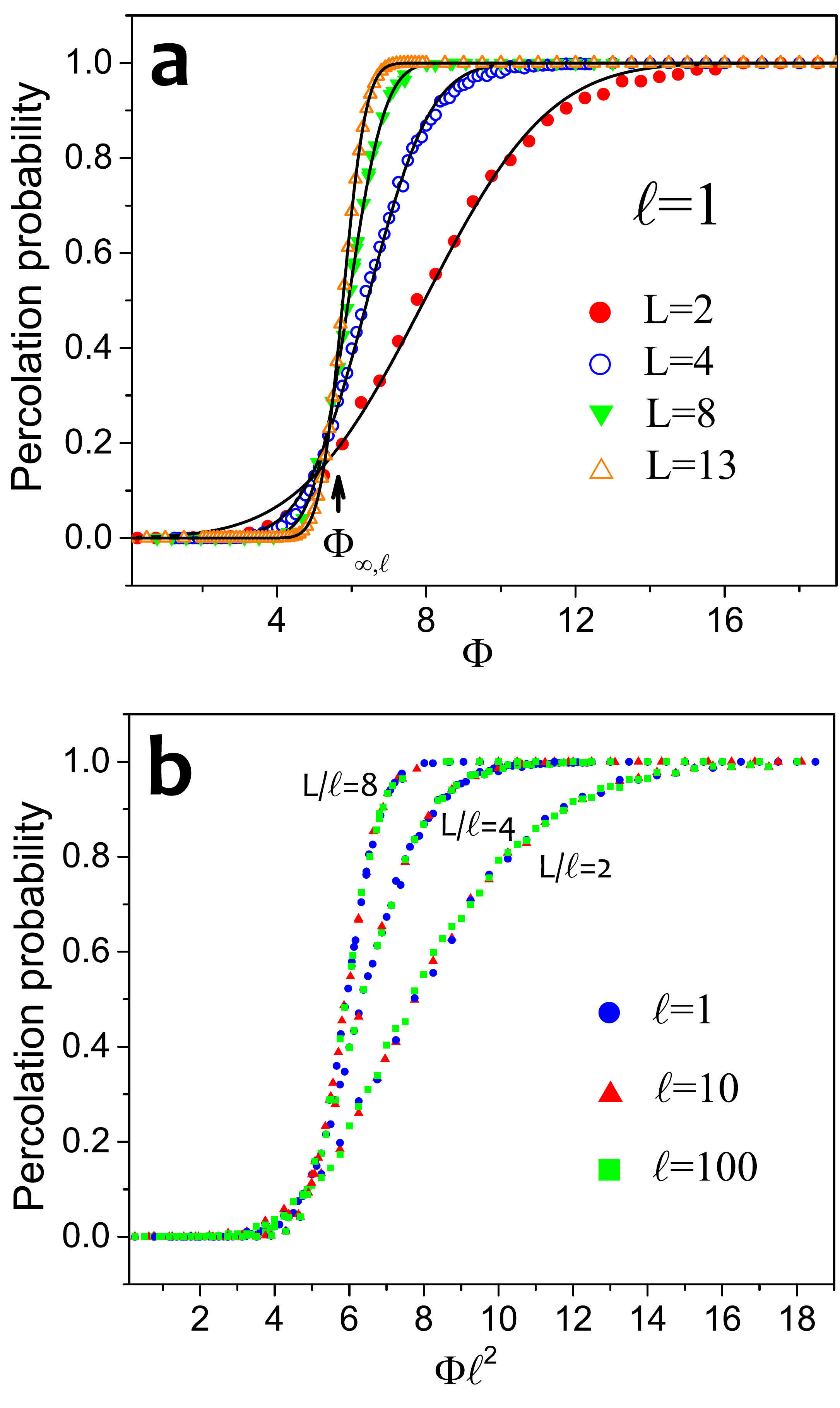}}
\caption{Percolation probability $\wp_{L,\ell}$
for square systems of percolating sticks with isotropic angular distribution.
(a) Stick length $\ell=1$;
the continuous-line fits assume a Gaussian distribution, Eq.\ \eqref{eq:erf}.
(b) Various values of $L$ and $\ell$, displaying the collapse of data
when the percolation probability is plotted against $\Phi\ell^2$.
}
\label{fig:P_vs_Phi}
\end{figure}

In order to validate our computer-simulation techniques we first 
reproduce some key scaling results available in the literature on 
stick percolation for square systems.
Let us consider square systems (aspect ratio $r=1$) of side $L$
with $\mathscr{N}$ sticks of fixed length $\ell$.
The stick density is thus $\Phi=\mathscr{N}/L^2$.
The percolation probability $\wp_{L,\ell}(\Phi)$ 
(i.e.\ the probability that there is at least
one spanning cluster) is shown in Figure \ref{fig:P_vs_Phi}(a) 
as a function of $\Phi$ for different values of $L$ and $\ell$.
The percolation probability is estimated for each density by running  
$n>1000$ realizations.

For finite systems, we wish to obtain the critical density
$\langle \Phi \rangle_{L,\ell}$, which, according to the 
Renormalization Group (RG) theory for square systems ($r=1$), 
scales with the system size $L$ as
\cite{sta-aha,hov-aha,zez-sta-bel,mer-mor,li-zha}
\begin{equation}
 \langle \Phi \rangle_{L,\ell} = \Phi_{\infty,\ell}+ 
   a_{\ell} \, L^{-\frac{1}{\nu}-\vartheta}.
\label{eq:scaling_square}
\end{equation}
The universal scaling exponent is $\nu=4/3$ for all two-dimensional 
percolation systems including lattice and continuum percolation 
(in particular, 2D random sticks systems). Recently it has been found 
that in random stick percolation square systems ($r=1$) the  non-analytical 
correction given by the exponent $\vartheta$ take the value $0.83\pm0.02$, 
consistent with previously published results \cite{zif-new,zif,aha-hov,sta}, 
while in rectangular systems ($r\neq1$) $\vartheta<0.83$ \cite{zez-sta-bel}.
From curves like the ones shown in Figure \ref{fig:P_vs_Phi}(a) one 
can extract
the size-dependent critical percolation density 
$\langle \Phi \rangle_{L,\ell}$ from the condition
\begin{equation}
 \wp_{L,\ell}(\langle \Phi \rangle_{L,\ell}) = \frac{1}{2},
 \label{eq:crit_dens_easy}
\end{equation}
which is satisfied if the probability distribution function 
\begin{equation}
 \Gamma_{L,\ell}(\Phi)= \frac{\partial\wp_{L,\ell}(\Phi)}
                             {\partial \Phi}
\end{equation}
%mer-mor
is usually assumed to be Gaussian \cite{rin-tor,zen-xu-she}.
A quick and rough estimate of the asymptotic critical percolation density 
$\Phi_{L\rightarrow\infty,\ell}$ (or $\Phi_{\infty,\ell}$) 
can be obtained by simulating a system with large $L$ and
employing Eq.\ (\ref{eq:crit_dens_easy}).
The curve with $L=13$ and $\ell=1$ in Figure \ref{fig:P_vs_Phi}(a) shows 
a fairly sharp transition at the critical stick density 
$\langle\Phi\rangle_{L=13,\ell=1}=5.754$, 
which is in fairly good agreement with the accepted value of 
$\Phi_{\infty,\ell=1}=5.6372858(6)$ \cite{mer-mor}.

A more accurate method to obtain 
$\langle \Phi \rangle_{L,\ell}$, which also allows us to obtain the
standard deviation of the distribution function, $\Delta_{L,\ell}$,
consists of fitting 
the simulated values of $\wp_{L,\ell}$ by a error-function erf($x$) 
according to Eq. (5)
\begin{equation}
 \wp_{L,\ell}(\Phi) = \frac{1}{2}
                              \left[1+\text{erf}
                                \left(
                                  \frac{\Phi-\langle \Phi \rangle_{L,\ell}}
                                       {\Delta_{L,\ell}}
                                \right)
                              \right],
\label{eq:erf}
\end{equation}
where $\text{erf}(x)=\frac{2}{\sqrt{\pi}}\int_0^x e^{-t^2} dt$.
Excellent fits (R$^2\geq0.9992$) validating the assumption of Gaussian distribution
are displayed in Figure \ref{fig:P_vs_Phi}(a) as solid lines.

\begin{figure} %3
\centerline{\includegraphics[width=10cm]{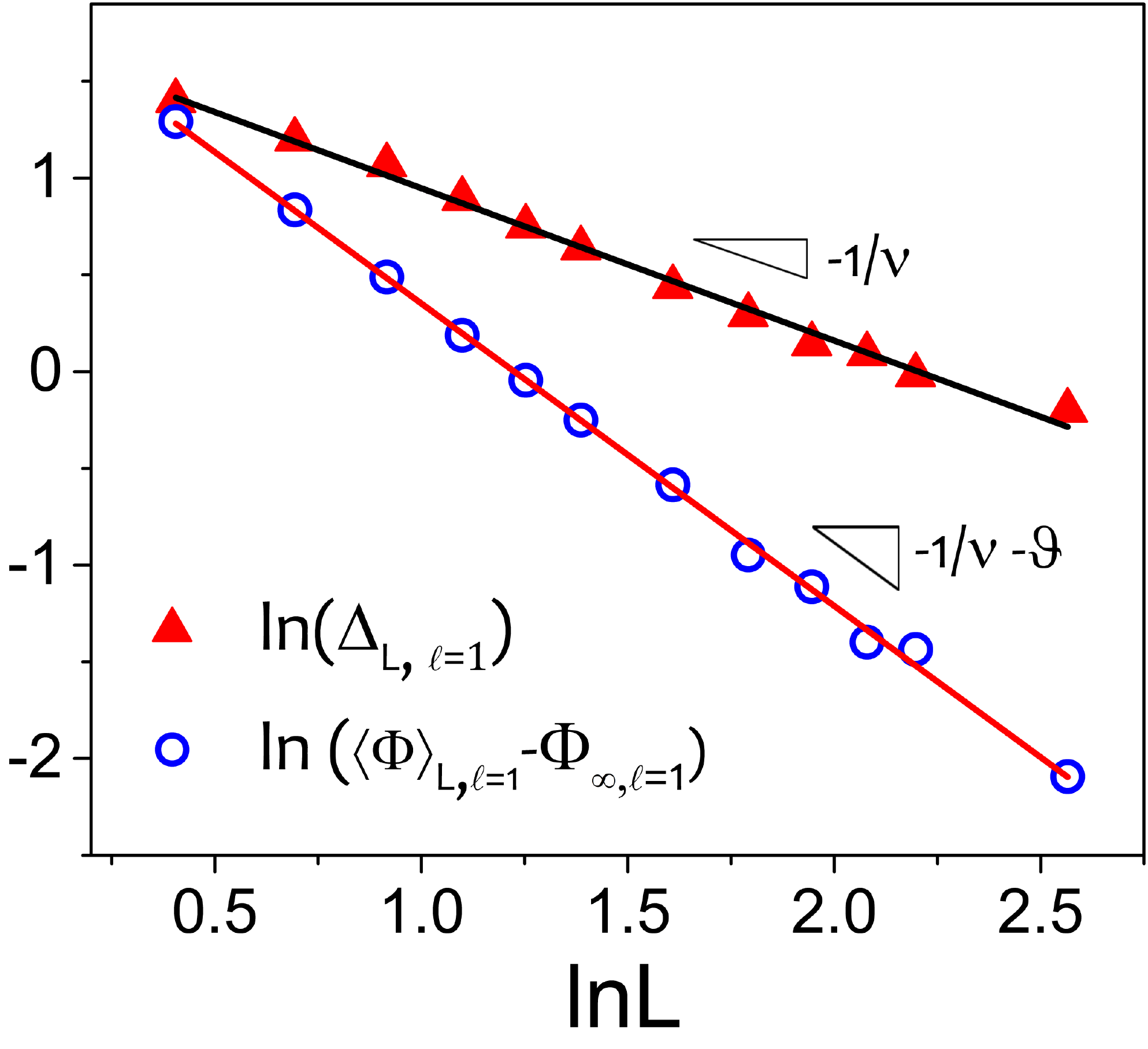}}
\caption{Critical percolation density $\langle \Phi \rangle_{L,\ell}$
and standard deviation $\Delta_{L,\ell}$
versus system size $L$ for square systems with stick length $\ell=1$ 
(measured in the same units as $L$).
%$\langle \Phi \rangle_{L,\ell}$ is extracted from the simulation data
%by assuming the accepted value of $\Phi_{\infty,\ell}=5.6372858(6)$
}
\label{fig:Phicrit_vs_L}
\end{figure}

As expected, in Figure \ref{fig:P_vs_Phi}(a) the percolation transition 
becomes less sharp for diminishing system size $L$ (according to the RG
theory the standard deviation scales as 
$\Delta_{L,\ell}\propto L^{-1/\nu}$ for isotropic square systems), 
and at the same time the critical density $\langle \Phi \rangle_{L,\ell}$ 
shifts towards higher values [Eq.\ \eqref{eq:scaling_square}].
Our simulations accurately reproduce the expected scaling laws for the
critical density and the standard deviation, as seen in 
Figure \ref{fig:Phicrit_vs_L} for stick length $\ell=1$.
From these curves we obtain $\vartheta=0.83\pm0.04$
and $\nu=1.33 \pm 0.03$
in agreement with previously reported values \cite{zif-new,zif,zez-sta-bel}.

The dependence of the percolation probability on the stick length, 
$\ell$, can be understood thanks to the RG theory, which suggests that
there is a relationship between $\Phi_{\infty,\ell}$ and $\ell$ given
by the effective area associated to each percolating element.
Associating an effective self-area $\ell^2$ to each stick, 
one obtains the relationship \cite{li-zha,mer-mor}
\begin{equation}
 \Phi_{\infty,\ell} \ell^2 = \Phi_{\infty,\ell=1}= 5.6372858(6).
\label{eq:phiellsq}
\end{equation}
Plotting the percolation probability for different values of $\ell$ 
as a function of $\Phi \ell^2$ (instead of doing it simply 
as a function of $\Phi$) shows that the most relevant quantity
is the ratio $L/\ell$ rather than $L$ and $\ell$ independently.
In simple terms, this indicates that what matters in the definition of
the density of sticks is the system size 
measured in units of the typical size of the percolating objects.
In Figure \ref{fig:P_vs_Phi}(b) we plot the percolation probability for
three values of $\ell=1, 10, 100$ and corresponding values of the
system size $L$ such that $L/\ell=2,4,8$.
The excellent collapse of curves with equal ratio $L/\ell$ ratifies
the validity of the RG analysis leading to Eq.\ \eqref{eq:phiellsq}.

%%%%%%%%%%%%%%%%%%%%%%%%%%%%%%%%%%%%%%%%%%%%%%%%%%%%%%%%%%%%%%%%%%%%%%%%%%%%%%%
\section{Synthesis and Experimental Morphological 
characterization of MRE}
\label{sec:experiments}

The elastomeric material that we studied is comprised of a 
polydimethylsiloxane (PDMS) polymer matrix and of percolating chains 
consisting of hybrid magnetite--silver microparticles.
These microparticles have an internal structure consisting of clusters of 
magnetite nanoparticles covered with metallic silver.
Using the hybrid filler material described above allows us to obtain 
a current--conducting magnetorheological material in superparamagnetic state 
(because magnetite nanoparticles, due to their small diameter, are in a 
superparamagnetic state at temperatures higher than their blocking temperature, 
$T_B =179 \, \text{K}$).
The electrical conductivity is not affected by oxidation, given by the chemical 
fastness of silver metal.
Finally, the use of PDMS as polymer matrix increases the chemical resistance 
of the composite against various chemical agents such as aromatic solvents, 
halogenated aliphatic solvents, aliphatic alcohols and concentrated salt 
solutions \cite{mie-jor-per,mie-rui-ant}.

The preparation method used to obtain the structured MRE composite with 
magnetic Fe$_3$O$_4$ silver-covered microparticles in PDMS (referred to as 
PDMS-Fe$_3$O$_4@$Ag) was described in detail in previous works 
\cite{mie-jor-per,mie-rui-ant} and is briefly described here.
First, Fe$_3$O$_4$ superparamagnetic nanoparticles (NPs) were synthesized by 
the chemical co--precipitation method where a solution mixture (2:1) 
of FeCl$_3 \cdot 6$ H$_2$O and FeCl$_2 \cdot$ 4H$_2$O in chlorhydric acid 
was added drop--by--drop to a solution of NaOH 
($60\,^{\circ}\mathrm{C}$, pH = 14), 
under nitrogen atmosphere and high--speed stirring.
The obtained nanocrystals were separated by repeated centrifugation 
and washing cycles, then dried in a vacuum oven at $40\,^{\circ}\mathrm{C}$
during 24 h.
The obtained dark brown NPs show a size distribution (determined by TEM 
images) with maximum at 13~nm in the log--normal distribution of diameters, 
which is in excellent agreement with the size of the crystallite domains 
calculated using the Debye--Scherrer relation from X--ray difractograms 
(XRD), ($14 \pm 2$) nm 
\cite{mie-jor-per,mie-rui-ant,god-mor-jor,but-alv-jor}.

In a second step, the Fe$_3$O$_4$ NPs were covered with silver in order 
to obtain electrically conductive and superparamagnetic particles.
For that, aqueous dispersions of Ag(NH$_3$)$_2^+$ and Fe$_3$O$_4$ NPs 
in a 10:1 molar ratio were sonicated for 30 min at room temperature. 
Then the system was heated in a water bath at  $40\,^{\circ}\mathrm{C}$
for 20 minutes with slow stirring.
In the following step, 0.4 M glucose monohydrate solution was added 
drop--by--drop to the Fe$_3$O$_4$--Ag$^+$ suspension.
Stirring was continued for one hour.
This synthesis protocol promotes the reduction of Ag (I) ions adsorbed 
onto Fe$_3$O$_4$ particles.
The magnetite--silver particles were separated out from the solution 
by magnetization and then by centrifugation.
After the particles were separated, the decanted supernatant liquid was 
fully transparent.
The obtained system (referred to here as Fe$_3$O$_4@$Ag) is actually formed by 
microparticles whose internal structure consists in several Fe$_3$O$_4$ 
nanoparticles clusters covered by metallic silver grouped together.
For the Fe$_3$O$_4@$Ag microparticles (MPs) the maximum of the diameter 
distribution is at 1.3 $\mu\text{m}$ (determined by SEM and TEM images).
For comparison purposes, silver particles (reddish orange) were produced 
in a separate batch using the same experimental conditions for each set.

Finally polidimethylsoloxane (PDMS) base and curing agent, referred to
as PDMS from now on (Sylgard 184, Dow Corning), were mixed in proportions 
of 10:1 (w/w) at room temperature and then loaded with the magnetic 
Fe$_3$O$_4$ silver-covered microparticles.
The amounts of PDMS and fillers were weighed during mixing on an 
analytical balance, homogenized and placed at room temperature in a vacuum 
oven for about two hours until the complete absence of any air bubble
is achieved. 
Specifically, composite material with 5\% w/w of Fe$_3$O$_4@$Ag was 
prepared. 
The still fluid samples were incorporated into a specially designed 
cylindrical mould (1 cm diameter by 1.5 cm thickness) and placed in 
between the magnetic poles of a Varian Low Impedance Electromagnet 
(model V3703), which provides highly homogeneous steady magnetic fields.
The mould was rotated at 30 rpm to preclude sedimentation and heated 
at $(75 \pm 5)\,^{\circ}\mathrm{C}$
in the presence of a uniform magnetic field 
($H_{\text{curing}}$= 0.35 T) during 3 hours to obtain the cured material.
The polymeric matrix is formed by a tridimensional crosslinked siloxane 
oligomers network with Si-CH$_2$-CH$_2$-Si linkages 
\cite{efi-wal-gen,est-bro-lav}.
Slices of the cured composites were hold in an ad-hoc sample--holder 
and cut using a sharp scalpel, which were used for the morphological 
(SEM and optical microscopy analysis) and electrical characterization 
of material. 

All fabricated composites obtained following the procedure described 
above displayed total electrical anisotropy, showing significant electrical 
conductivity only in the direction of application of the magnetic 
field during curing (which coincides with the direction of 
Fe$_3$O$_4@$Ag chains). 
For these MRE materials, electrical resistivity values of
$\rho_{\parallel}=4\, \Omega \times \text{cm}$ 
and $\rho_{\perp}=60\, \text{M}\Omega \times \text{cm}$
were obtained, where $\parallel$ and $\perp$ indicates parallel 
and perpendicular direction with respect to the filler needles, respectively.

Then, we proceeded to the morphological characterization, 
evaluating the angular, length, and diameter distributions of 
the conductive chains.
This analysis was performed by computing the angle, length, and diameter 
of the Fe$_3$O$_4@$Ag chains in SEM images to several zooms (50x to 6000x) 
and images obtained by optical microscopy using the image processing 
software ImageJ v1.47. 
For SEM images, voltage (EHT) 5 KV and extensions of 100x (3300 pixels.cm-1) 
and 300x (9800 pixels.cm-1) were typical conditions to compute the 
average chain length, while to compute chain diameters 3 KV voltages 
and 4000x (40 pixels.$\mu$m-1) were used. 

\begin{figure} %4
\centerline{\includegraphics[width=10cm]{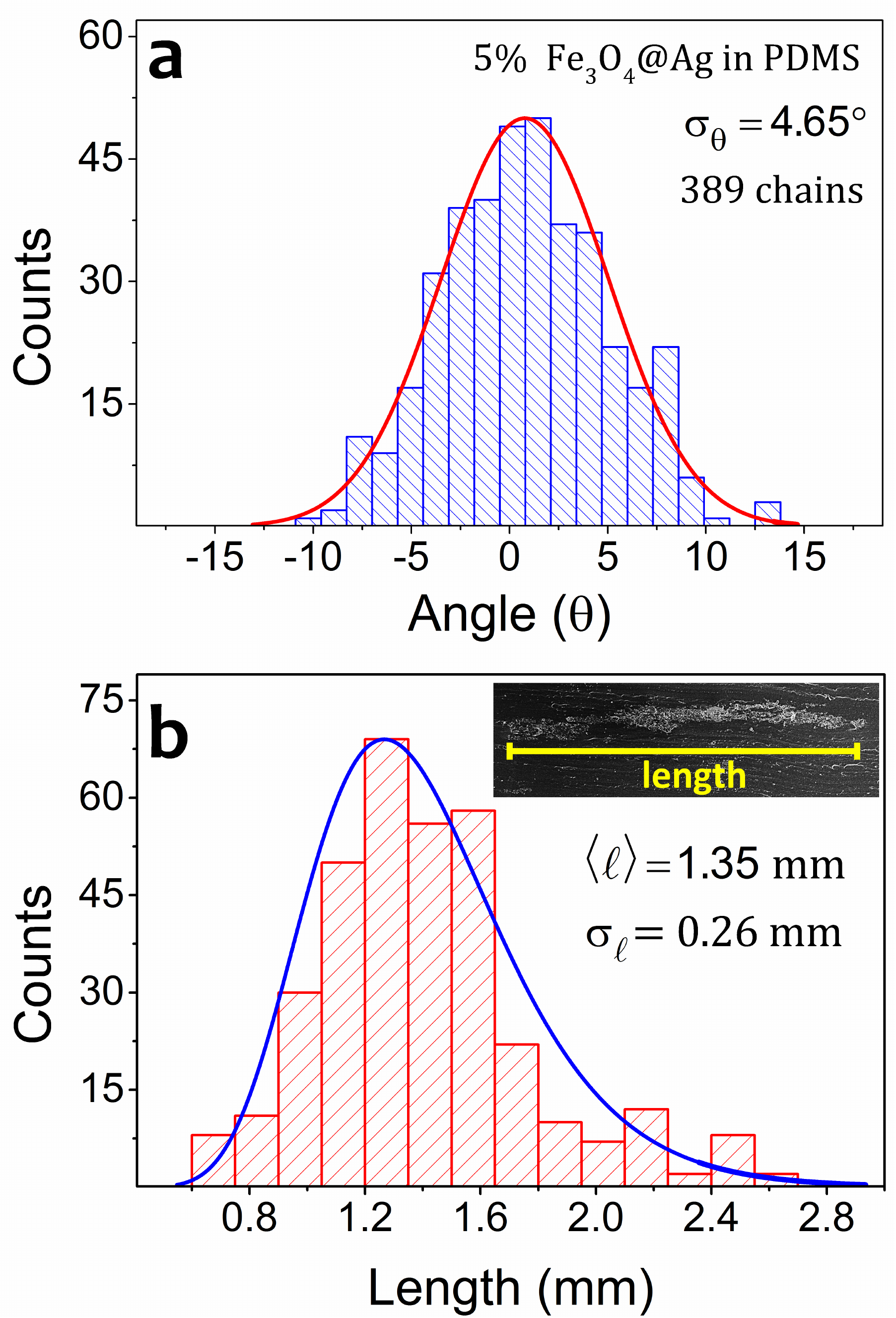}}
\caption{(a) Histogram for the angular distribution of chains
in the MRE PDMS-Fe$_3$O$_4@$Ag 5\% w/w.
The histogram is adjusted by a Gaussian distribution function (solid line).
(b) Histogram associated with the distribution of 
chain lengths, built by measuring the length of 364 chains.
The histogram is adjusted by a log-normal distribution 
function (solid line).
Inset: SEM image of a chain.}
\label{fig:distribucion_MRE}
\end{figure}

Figure \ref{fig:distribucion_MRE}(a) shows the histogram obtained for the 
angular distribution of the chains.
The histogram is adjusted by a Gaussian distribution function (solid line)
centered on the direction of application of the magnetic field during 
curing $H_{\text{curing}}$ 
($\theta=0$) with standard deviation $\sigma_{\theta}= (4.65\pm0.02) \,^{\circ}$ 
(count performed on 389 chains).

Figure \ref{fig:distribucion_MRE}(b) shows the histogram associated with 
the distribution 
of chain lengths in the MRE PDMS-Fe$_3$O$_4@$Ag 5\% w/w, built by 
computing 364 chain lengths.
It has an excellent degree of adjustment ($R^2=0.9965$) with 
the log-normal distribution function
\begin{equation}
 f_\lambda=p\left(\lambda;
                  \langle \lambda \rangle, 
                  \sigma_\lambda
            \right)
          =\frac{1}{\sqrt{2\pi} \sigma_\lambda \lambda}
            \exp{
                 \left(\frac{-[\ln\lambda-\ln\langle\lambda\rangle]^2}
                           {2\sigma_{\lambda}^2}
                 \right)
                }
\label{fig:flambda}
\end{equation}
with $\lambda=\ell$ and fitted parameters 
$\langle\ell\rangle =(1.35\pm0.01) \, \text{mm}$ y
$\sigma_\ell= (0.26\pm0.01) \, \text{mm}$.
The average stick density observed is 11.84 needles$/$mm$^{2}$.
Note that here, as we start to study concrete physical systems,
we begin to use regular units of length, such as millimeters.

The histogram of values of diameters (not shown) was built from 311 counts 
and also adjusted by a log-normal distribution [Eq.\ \eqref{fig:flambda} 
with $\lambda=d$]
with an excellent degree of adjustment, $R^2 = 0.9977$, and 
parameters $\langle d\rangle =(10.40\pm0.02) \, \mu$m and 
$\sigma_d =(0.30\pm0.01) \, \mu$m.
Note that, under the used experimental conditions, the length
of the chain is much greater than its diameter, so that the 
approximation of one-dimensional percolating elements is justified.

%%%%%%%%%%%%%%%%%%%%%%%%%%%%%%%%%%%%%%%%%%%%%%%%%%%%%%%%%%%%%%%%%%%%%%%%%%%%%%%
\section{Simulation of Anisotropic Systems}

From the point of view of possible technological applications of MRE,
it is important to characterize the degree of anisotropy of the electrical 
conductivity of a given device.
Anisotropy can be introduced essentially in two ways: 
through an aspect ratio $r\neq 1$ which makes the system
asymmetric (relevant when the characteristic length
of the percolating objects is not much smaller than the size of the 
system, as in our study), and through an anisotropic angular distribution
of the sticks.
These two aspects can be controlled experimentally in systems like the ones
discussed in the previous section, and we will now incorporate them in our
simulation studies.

Let us denote the probabilities of having spanning clusters
as follows:
horizontally $\wp^{H}$, 
vertically $\wp^{V}$, 
only horizontally $\wp^{HX}$, 
only vertically $\wp^{VX}$, 
on either direction $\wp^{\,U}$, 
and on both directions $\wp^{HV}$. 
These probabilities are not independent of each other, as 
they satisfy $\wp^{\,U}=\wp^{H}+\wp^{V}-\wp^{HV}$ and 
$\wp^{\,U}=\wp^{HX}+\wp^{VX}+\wp^{HV}$ \cite{mer-mor}.
The signature of the presence of total anisotropy in the percolation
regime will be given by $\wp^{HX}$ taking values very close to 1.
Experimentally, a set of parameters that ensure that condition
will constitute what we can call a ``safety zone'' of total anistropic 
percolation and, therefore, of TEA. 
The main goal of the present work is to establish
a methodology and, with it, to arrive at
the specification of such a set of parameter values, as
an aid to obtaining TEA in devices that demand it
(pressure mapping sensors, 
bidimensional Zebra\textsuperscript{\textregistered}-like 
connectors, etc.).

\begin{figure} %5
\centerline{\includegraphics[width=10cm]{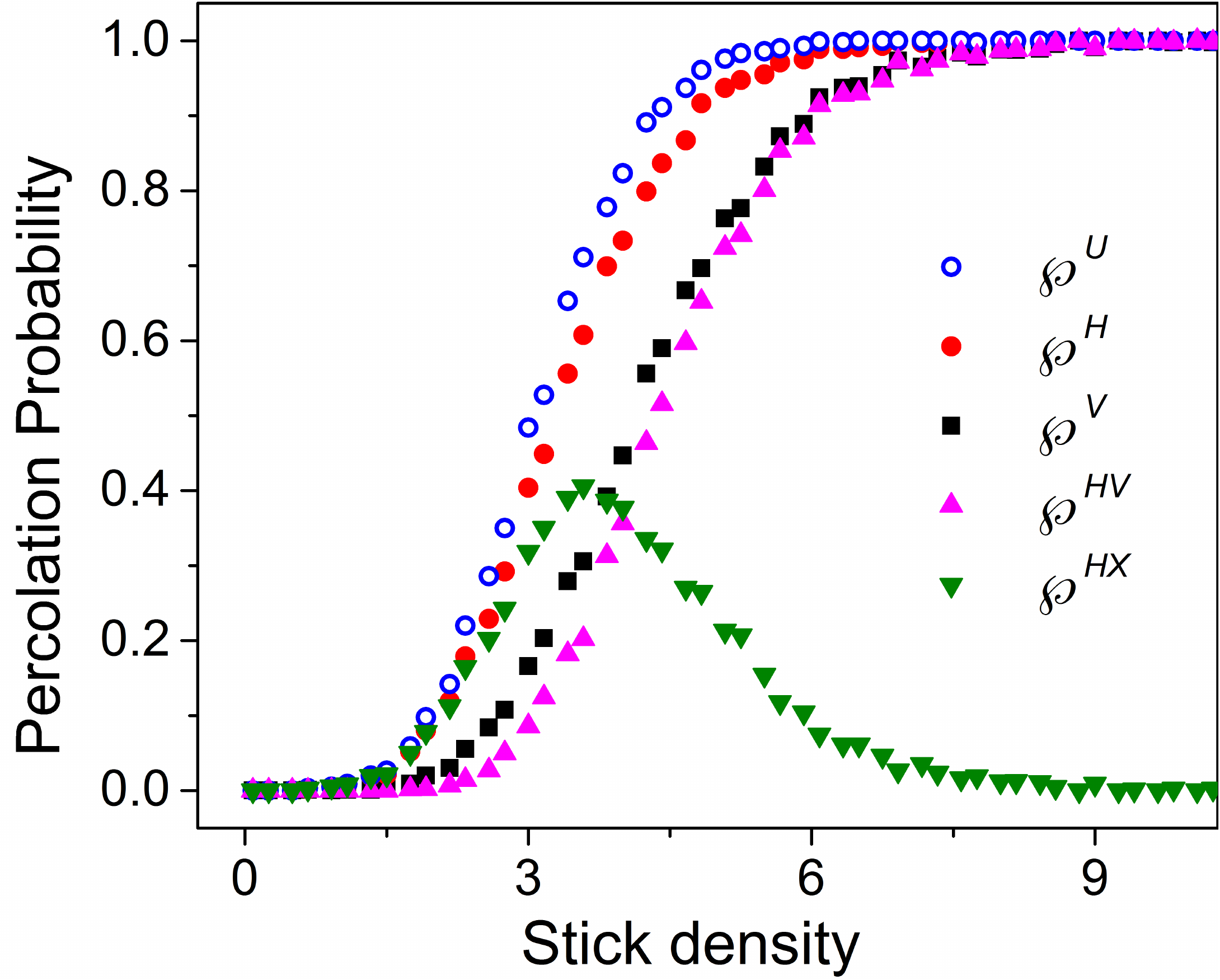}}
\caption{Various percolation probabilities 
($\wp^{H}, \wp^{V}, \wp^{HX}, \wp^{\,U}, \wp^{HV}$)
versus density of sticks ($\text{chains/mm}^2$) for
a rectangular system of aspect ratio $r=L_x/L_y=3/4$,
$L_x=3\,\text{mm}$, and isotropic stick distribution 
with $\langle\ell\rangle=1.35\,\text{mm}$ and 
$\sigma_\ell = 0.26\, \text{mm}$, for a log-normal distribution
of stick lengths.
Note that $\wp^{VX}$ is not shown since it is always negligible.
}
\label{fig:perc_prob_asym_iso_H_V}
\end{figure}

It is important to first determine whether the asymmetry of the box 
or the anisotropy of the angular distribution contribute equally 
or not to the global anisotropy of the percolation behavior.
In Figure \ref{fig:perc_prob_asym_iso_H_V} we show the percolation 
probabilities $\wp^{H}$, $\wp^{V}$, $\wp^{HX}$, $\wp^{\,U}$, and 
$\wp^{HV}$
for a rectangular isotropic (the stick angular distribution is
uniform, with $-\pi \le \theta \le \pi$) system with aspect ratio $r=3/4$,
$L_x=3  \, \text{mm}$, and log-normal length distribution parameters
$\langle \ell \rangle = 1.35 \, \text{mm}$, and
$\sigma_{\ell}=0.26 \, \text{mm}$
(notice that $\wp^{VX}$ is negligible and does not need to be
considered in the analysis).
We remark that these parameters are taken from an experimental
sample, as discussed above (Figure \ref{fig:distribucion_MRE}).
The different percolation probabilities verify the expected
inequalities  
$\wp^{\,U} \ge \wp^{H} \ge \wp^{V}\ge \wp^{HV}$,
given the chosen asymmetry of the box.
The values of $\wp^{HX}$ seen in this figure, never close to unity,
indicate that the mere asymmetry of the box ($r\neq 1$) is not
enough to produce a safety zone of totally anisotropic conduction.
Therefore, we conclude that in order to achieve effective TEA
in bulk or films sample geometries it is required to introduce an 
internal anisotropy, that is, in the stick angular distribution.
This conclusion is consistent with experimental observations
\cite{mie-jor-per,mie-rui-ant}.

We need to introduce a magnitude to characterize in a generic 
and quantitative way the degree of internal anisotropy of 
the system of random sticks.
Let us denote it macroscopic anisotropy and it will be given by
\begin{equation}
\mathscr{A}=\frac{\sum_{j=1}^{\mathscr{N}}\ell_j \left |\cos\theta_j \right|}
                 {\sum_{j=1}^{\mathscr{N}}\ell_j \left |\sin\theta_j \right|}.
\end{equation}
In the limit of infinite percolating objects
we have $\mathscr{A}=1$ for isotropic systems, 
while $\mathscr{A}\rightarrow\infty (0)$ for completely anisotropic 
systems favoring the horizontal (vertical) direction.

\subsection{Influence of $\sigma_\theta$}

\begin{figure} %6
\centerline{\includegraphics[width=15cm]{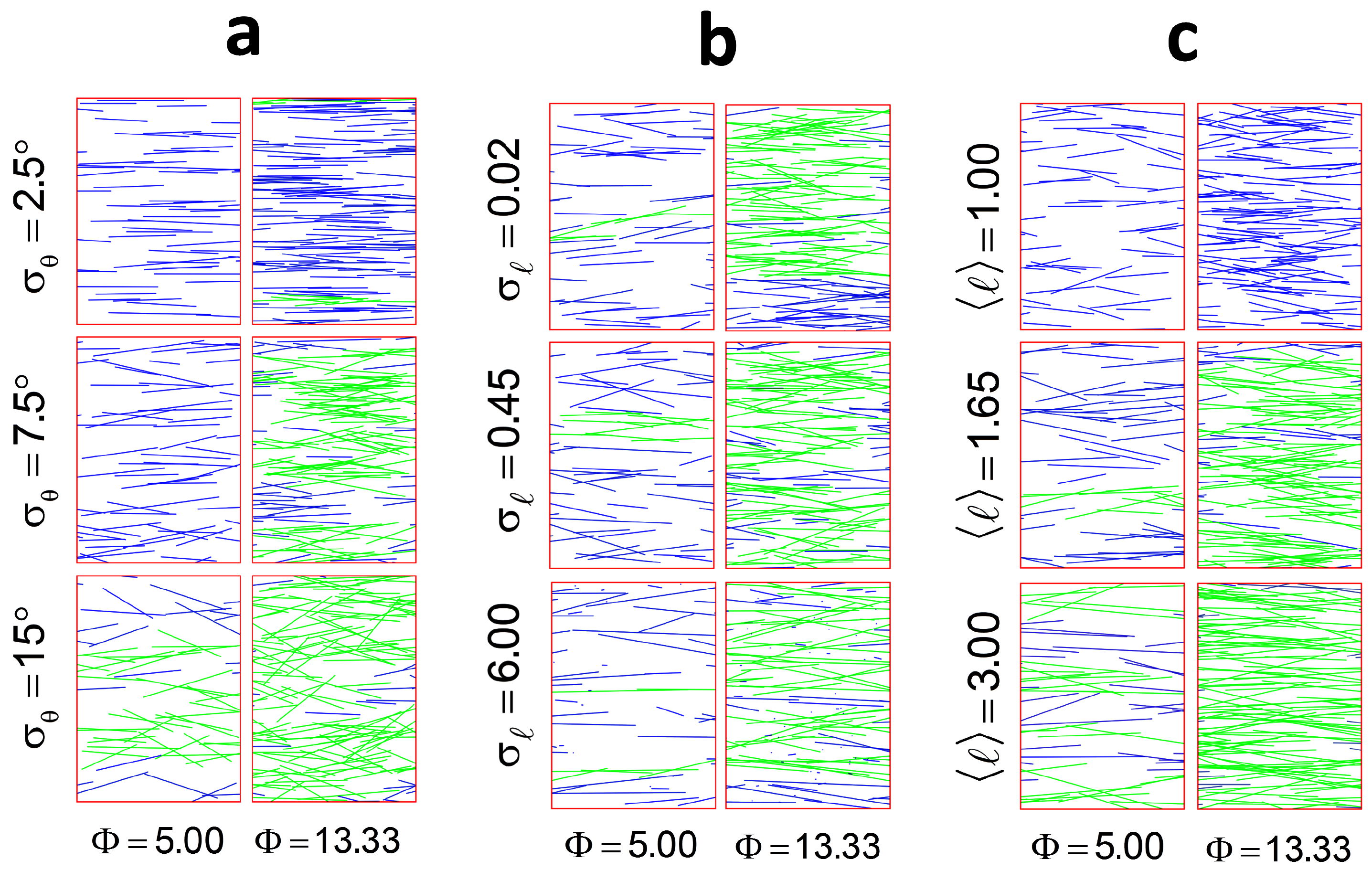}}
\caption{Examples of random stick systems in a two-dimensional box
of sides $L_x=3\,\text{mm}$ and $L_y=4\,\text{mm}$ with anisotropic 
angular distributions and non-uniform stick length for two different 
stick densities: 
$\Phi=5.00\, \text{chains/mm}^2$ and $\Phi=13.33\, \text{chains/mm}^2$.
The green (color online) sticks belong to a horizontal spanning cluster and the 
blue ones do not. 
The angle distribution is Gaussian centered in zero, and the length distribution
is log-normal (as found experimentally).
(a) $\langle\ell\rangle=1.35 \,\text{mm}$,
$\sigma_\ell=0.26 \,\text{mm}$,
(b) $\langle\ell\rangle=1.35 \,\text{mm}$, $\sigma_\theta= 7.5^{\circ}$
(c) $\sigma_\theta= 7.5^{\circ}$, $\sigma_\ell=0.26 \,\text{mm}$.
}
\label{fig:image_with_anisotropy}
\end{figure}

Figure \ref{fig:image_with_anisotropy} shows examples of random 
stick systems in a two-dimensional box of sides 
$L_x=3\,\text{mm}$ and $L_y=4\,\text{mm}$ (corresponding to the characteristic 
dimensions of the experimental samples)
with anisotropic angular distributions and non-uniform 
stick length. In all cases the green sticks belong to a horizontal spanning 
cluster and the blue ones do not.
In particular, Figure \ref{fig:image_with_anisotropy}(a) shows systems 
for three
different values of the standard deviation of the Gaussian angular 
distribution, $\sigma_\theta$, and two values of the stick density $\Phi$,
with parameters of a log-normal distribution
$\langle\ell\rangle=1.35 \,\text{mm}$ and 
$\sigma_\ell=0.26 \,\text{mm}$.
As expected, for a given value of $\sigma_\theta$, more sticks participate
in the spanning cluster the higher the density of sticks.
We also note that, for a fixed value of the density $\Phi$, the fraction 
of sticks that belong to the spanning cluster also increases 
with $\sigma_\theta$.

\begin{figure} %7
\centerline{\includegraphics[width=9cm]{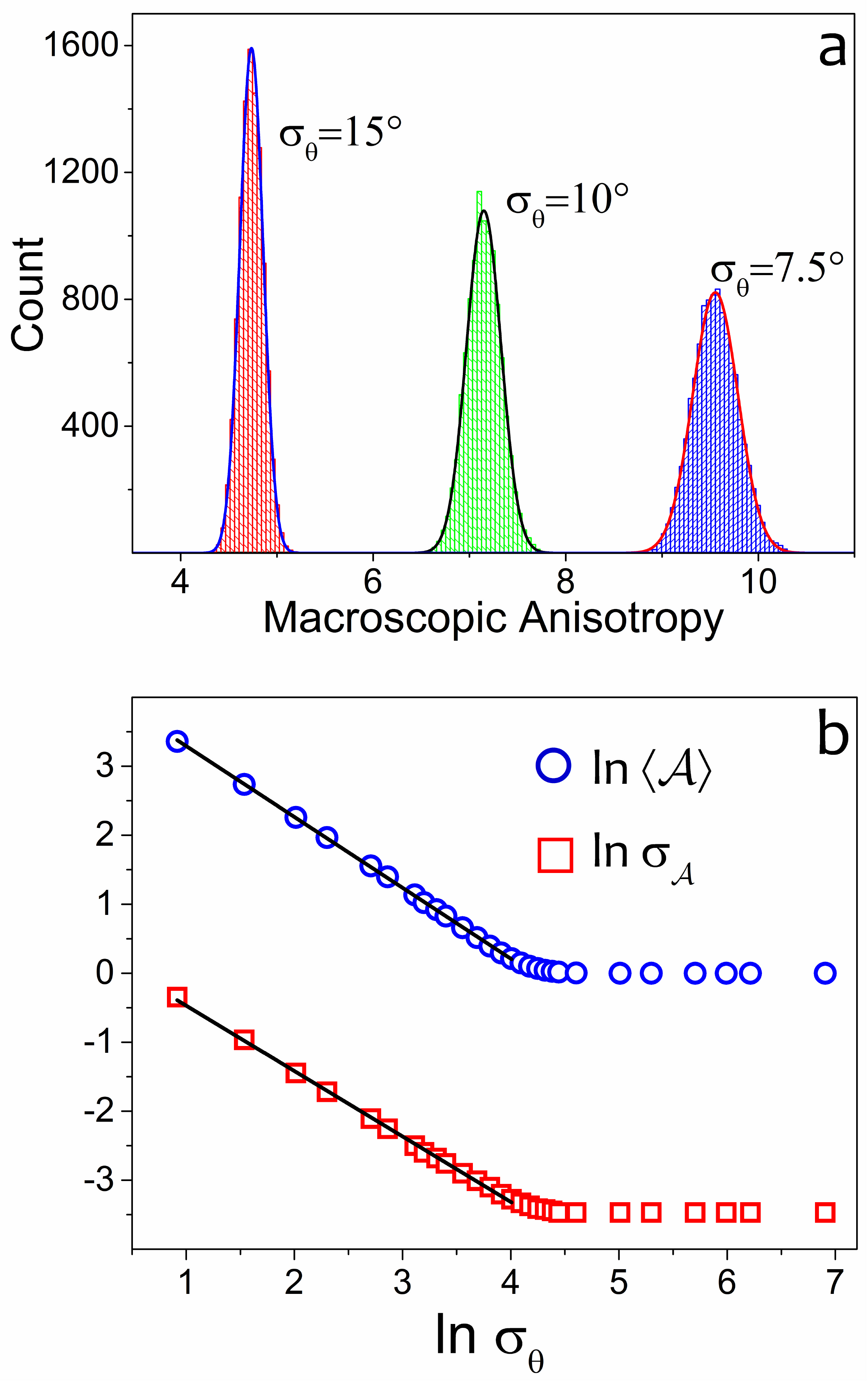}}
\caption{(a) Histograms of the macroscopic anisotropy $\mathscr{A}$ obtained 
for three different values of $\sigma_\theta$ and $\mathscr{N}=1000$ for a
rectangular system of aspect ratio 
$r=L_x/L_y=3/4$ and anisotropic stick distribution, with parameters 
of a log-normal distribution
$\langle\ell\rangle=1.35 \,\text{mm}$ and 
$\sigma_\ell=0.26 \,\text{mm}$.
(b) Macroscopic anisotropy and its standard deviation
versus $\sigma_\theta$.
For small $\sigma_\theta$ we obtain a scaling behavior with exponent
approximately equal to $-1$ for both quantities.}
\label{fig:anisotropia_macroscopica_sigma_theta}
\end{figure}

To evaluate the effect of $\sigma_\theta$, 
$\langle\ell\rangle$, and $\sigma_\ell$ on the TEA (i.e.\ on the formation of 
a spanning cluster only in 
the horizontal direction) numerical simulations of rectangular 
systems with $L_x = 3 \,\text{mm}$ and $L_y = 4 \, \text{mm}$ 
($r = 3/4$) were made, taking the horizontal direction ($\theta=0^\circ$) 
as the direction of application of the magnetic field during curing 
($H_{\text{curing}}$). 
In particular, to evaluate the effect of $\sigma_\theta$, 
a log-normal distribution for the lengths with parameters 
$\langle\ell\rangle=1.35 \,\text{mm}$ and 
$\sigma_\ell=0.26 \,\text{mm}$
(empirical parameters for the MRE PDMS-Fe$_3$O$_4$@Ag 5\% w/w), 
and a Gaussian angular distribution with parameters 
$\langle\theta\rangle=0^\circ$ and 
different values of standard deviation, $\sigma_\theta$, were used in
our simulations.
Figure \ref{fig:anisotropia_macroscopica_sigma_theta}(a) shows histograms of 
macroscopic anisotropy, $\mathscr{A}$ obtained for three different 
values of $\sigma_\theta$ ($15^\circ$, $10^\circ$, and $7.5^\circ$),  
each one obtained by performing 10500 repetitions, with
$\mathscr{N}=1000$.
For all values of $\sigma_\theta$, the distribution is approximately 
Gaussian, with an excellent degree of fitting, $R^2 \ge 0.99926$ 
[continuous-line in Figure \ref{fig:anisotropia_macroscopica_sigma_theta}(a)].
For these distributions, the average macroscopic anisotropy,
$\langle\mathscr{A}\rangle$, 
and its standard deviation, $\sigma_\mathscr{A}$, 
follow a monotonously decreasing behavior with $\sigma_\theta$, 
as illustrated in Figure \ref{fig:anisotropia_macroscopica_sigma_theta}(b). 
It is noteworthy that for small values of $\sigma_\theta$ 
($\sigma_\theta < 55^\circ$) there exists 
a linear relationship between 
$\ln \langle\mathscr{A}\rangle$ and $\ln \sigma_\theta$, 
as well as between $\ln \sigma_\mathscr{A}$ and $\ln \sigma_\theta$
[solid lines in Figure \ref{fig:anisotropia_macroscopica_sigma_theta}(b), 
with $R^2=0.9998$, slope = -1.02(7) and intercept= 4.32(1) 
for $\ln \langle\mathscr{A}\rangle$, 
and $R^2=0.99897$,  slope = -0.97(6), and intercept = 0.48(3) for 
$\ln \sigma_\mathscr{A}$].

As described above, a strategy to study the influence of $\sigma_\theta$
on the TEA of the composite material is to evaluate 
curves of $\wp^{HX}(\Phi)$ for different values of $\sigma_\theta$. 
Values of $\Phi$ for which $\wp^{HX}(\Phi)=1$  (if they exist) 
constitutes 'safety zones' in terms of TEA: 
for fixed values of 
$\langle\ell\rangle$, $\sigma_\ell$, $\sigma_\theta$, 
and densities of percolating objects $\Phi$
in this 'safety zone', systems are most likely to have TEA, 
i.e.\ electrical conductivity only in the horizontal direction 
by formation of a spanning cluster only in that direction.

\begin{figure} %8
\centerline{\includegraphics[width=15cm]{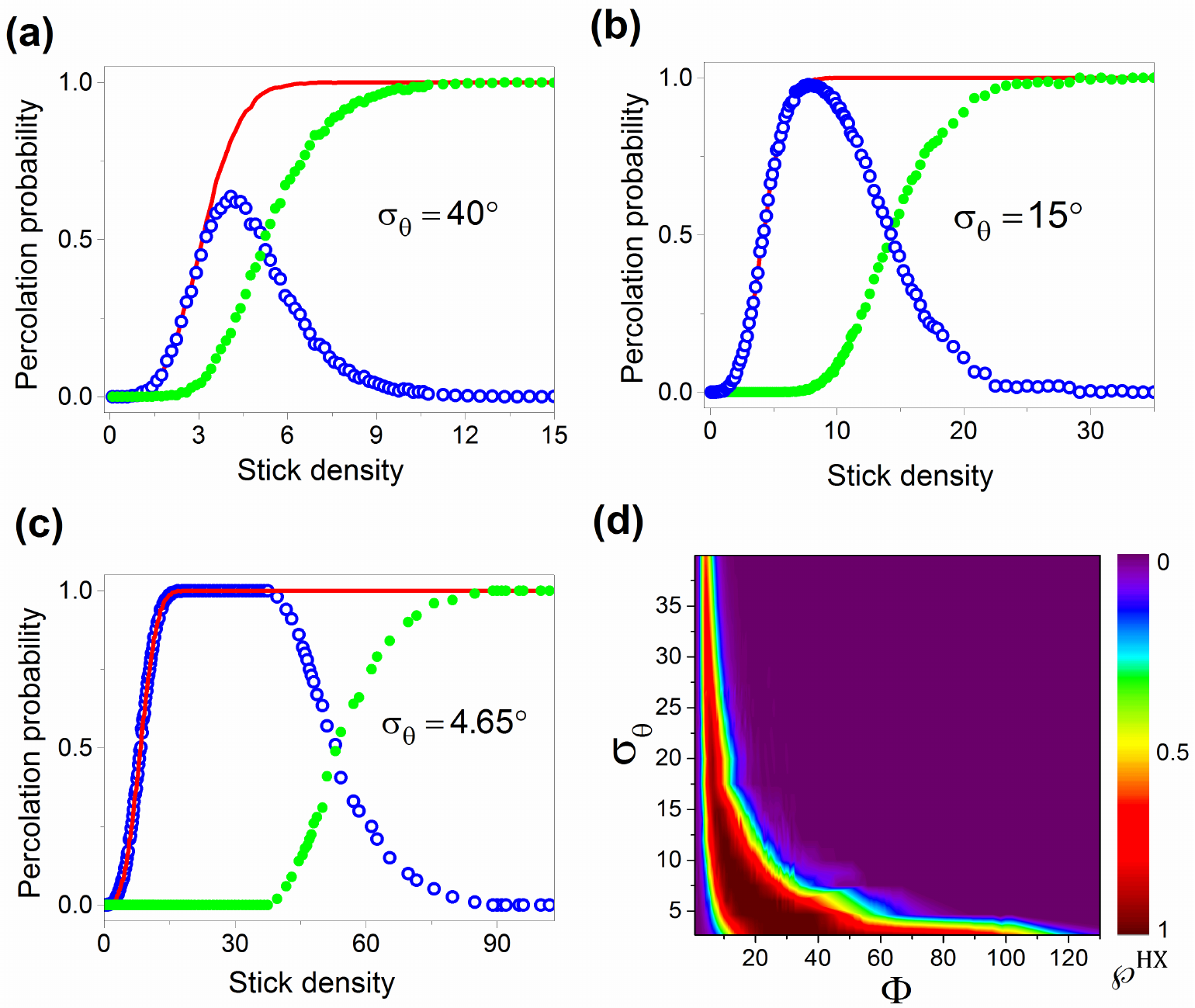}}
\caption{Various percolation probabilities 
($\wp^{H}, \wp^{V}, \wp^{HX}$)
versus density of sticks for a rectangular system of aspect ratio 
$r=L_x/L_y=3/4$ and anisotropic stick distribution, with parameters 
of a log-normal distribution
$\langle\ell\rangle=1.35 \,\text{mm}$ and 
$\sigma_\ell=0.26 \,\text{mm}$.
(a)-(c) Three different values of the standard deviation $\sigma_\theta$ of the 
angular Gaussian distribution. 
Red solid line: $\wp^{H}$, 
green filled circles: $\wp^{V}$,
blue open circles: $\wp^{HX}$.
(d) Probability $\wp^{HX}$ as a contour plot versus the stick density
$\Phi$ and $\sigma_\theta$, showing the full dependence on 
$\sigma_\theta$ not seen in the other panels.}
\label{fig:percolation_sigma_theta}
\end{figure}

Figure \ref{fig:percolation_sigma_theta}(a-c) show the curves 
of $\wp^{H}, \wp^{V}$, and $\wp^{HX}$,  
for three values of $\sigma_\theta$ 
($40^\circ$, $15^\circ$ and $4.65^\circ$) 
for systems with log-normal distribution for the stick lengths 
with parameters
$\langle\ell\rangle=1.35 \,\text{mm}$ and 
$\sigma_\ell=0.26 \,\text{mm}$. 
There are values of $\Phi$ for which $\wp^{HX}(\Phi)=1$ only 
when $\sigma_\theta < 15^\circ$.  
Such behavior of $\wp^{HX}(\Phi)$ is detailed in 
Figure \ref{fig:percolation_sigma_theta}(d), which shows the probability 
$\wp^{HX}(\Phi)$ as a contour plot of density versus $\Phi$ 
and $\sigma_\theta$.
It can be seen that the range of values of $\Phi$ for which 
$\wp^{HX}(\Phi)=1$ strongly increases with decreasing $\sigma_\theta$ 
and, also, with lower values of the parameter $\sigma_\theta$ higher 
stick density $\Phi$ is required to reach the safety zone.

As described in Section \ref{sec:experiments}, 
all the PDMS-Fe$_3$O$_4$@Ag 5\%w/w 
systems synthesized have $\Phi = 11.84\, \text{chains/mm}^2$ and 
electrical anisotropy (measurable electrical conductivity 
only in the direction of the magnetic field applied  
during the curing of the material). 
Panel (c) of Figure \ref{fig:percolation_sigma_theta} shows that, 
for this value of $\Phi$ and the parameters $\sigma_\theta=4.65^\circ$,  
$\langle\ell\rangle=1.36 \,\text{mm}$ and 
$\sigma_\ell=0.26 \,\text{mm}$
(experimental parameters for PDMS-Fe$_3$O$_4$@Ag 5\%w/w), 
we have a very high only-horizontally percolation probability, 
which shows a very good correlation between our performed 
simulations and the experimental results obtained.

%%%%%%%%%%%%%%%%%%%%%%%%%%%%%%%%%%%%%%%%%%%%%%%%%%%%%%%%%%%%%%%%%%%%%%%%%%%%%%%
\subsection{Influence of $\sigma_\ell$}

Following a similar procedure to the one described in the previous section, 
in order to evaluate the effect of $\sigma_\ell$, a Gaussian angular 
distribution with parameters $\langle\theta\rangle=0^\circ$ and 
$\sigma_\theta=7.5^\circ$, and a log-normal distribution 
for the lengths with parameters 
$\langle\ell\rangle=1.35\,\text{mm}$ and different values 
of $\sigma_\ell$ were used in our simulations.
Figure \ref{fig:image_with_anisotropy}(b) shows systems for three
different values of the standard deviation of the log-normal 
length distribution, $\sigma_\ell$, and two values of the stick density $\Phi$,
with structural parameters $\langle\ell\rangle=1.35 \,\text{mm}$ and 
$\sigma_\theta=7.5^\circ$.

\begin{figure} %9
\centerline{\includegraphics[width=15cm]{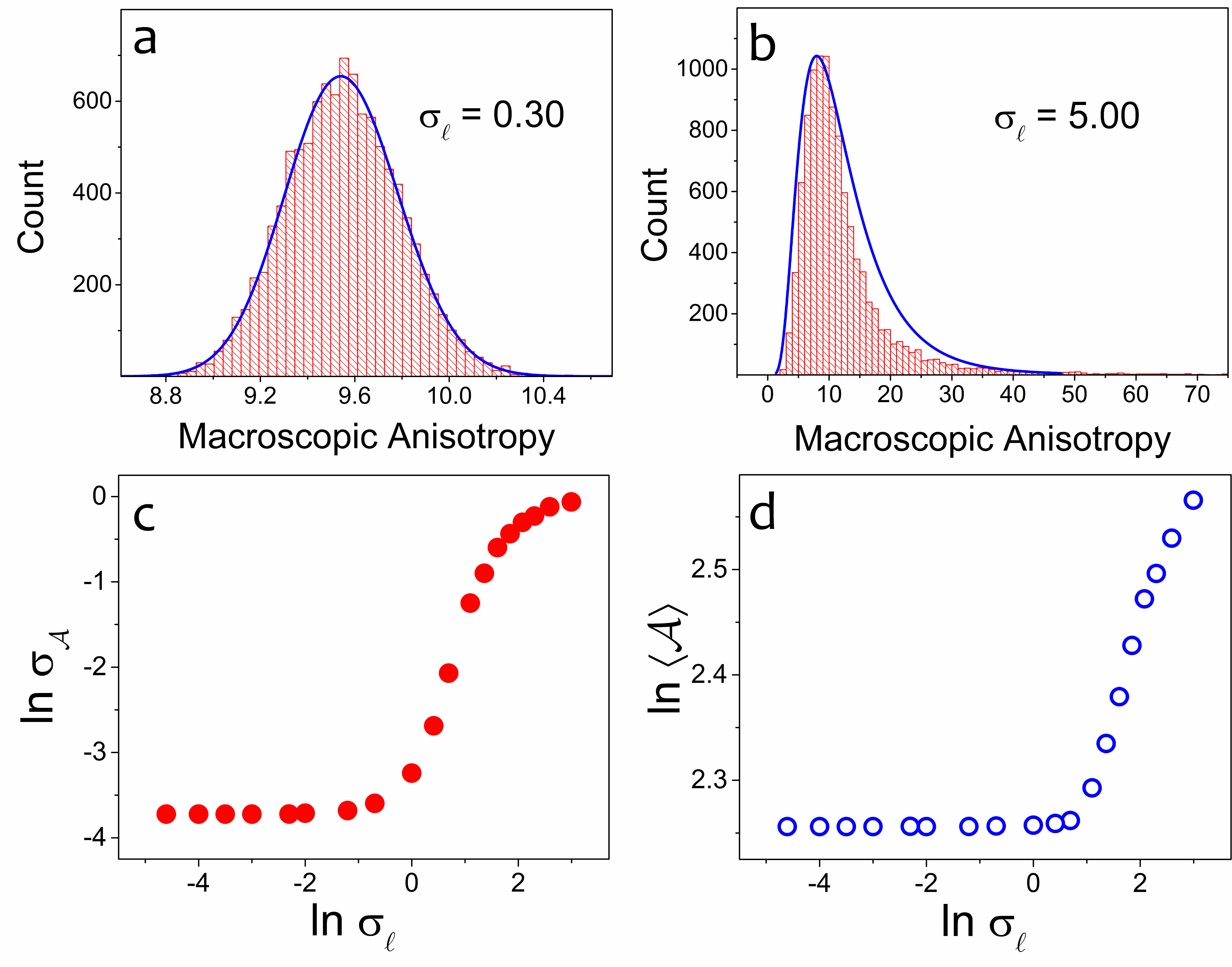}}
\caption{(a,b) Histograms of the macroscopic anisotropy $\mathscr{A}$ 
obtained for two different values of $\sigma_\ell$ and $\mathscr{N}=1000$
for a rectangular system of aspect ratio 
$r=L_x/L_y=3/4$ and parameters
$\langle\ell\rangle=1.35 \,\text{mm}$ and $\sigma_\theta= 7.5^{\circ}$.
(c,d) Macroscopic anisotropy and its standard deviation
versus $\sigma_\ell$.
}
\label{fig:anisotropia_macroscopica_sigma_length}
\end{figure}

\begin{figure} %10
\centerline{\includegraphics[width=8cm]{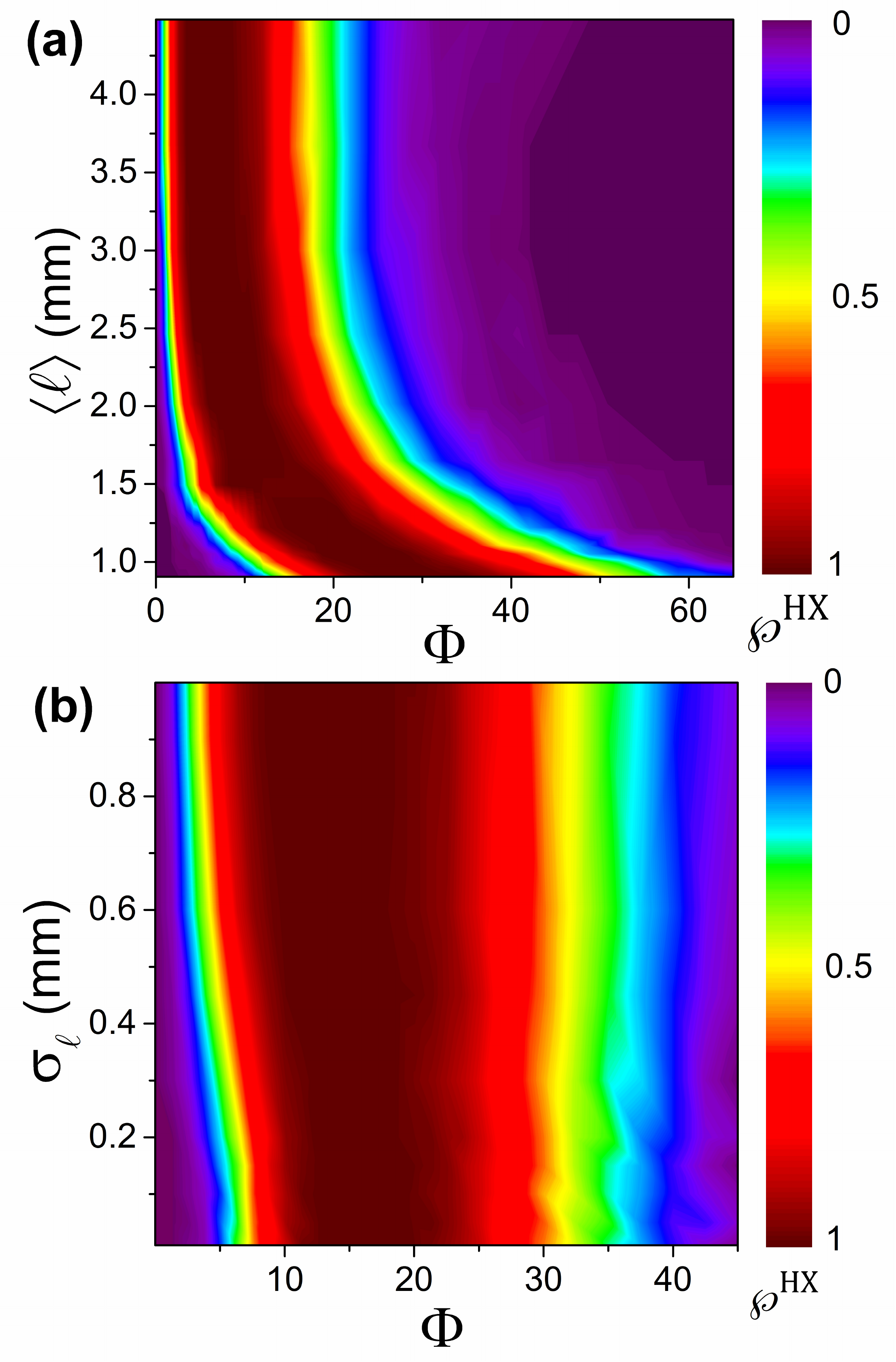}}
\caption{Probability $\wp^{HX}$ as a contour plot versus the stick 
         density $\Phi$ and
(a) $\langle\ell\rangle$ with
    $\sigma_\theta= 7.5^{\circ}$ and $\sigma_\ell=0.26 \,\text{mm}$
(b) $\sigma_\ell$ with 
    $\langle\ell\rangle=1.35 \,\text{mm}$ and $\sigma_\theta= 7.5^{\circ}$.}
\label{fig:contour_plot_ell_sigmaell}
\end{figure}

Figure \ref{fig:anisotropia_macroscopica_sigma_length} shows histograms 
of macroscopic anisotropy $\mathscr{A}$ obtained 
for two different values of $\sigma_\ell$ (0.30 mm  and 5.00 mm), 
each one obtained by performing 10500 repetitions, 
with $\mathscr{N}=1000$.  
For all the values of $\sigma_\ell$, the distribution is approximately 
log-normal, with an excellent degree of adjustment, $R^2\geq0.99511$ 
[continuous-line in
Figure \ref{fig:anisotropia_macroscopica_sigma_length}(a-b)].
At low values of $\sigma_\ell$ the distribution is approximately Gaussian.
For these distribution, the average macroscopic anisotropy,
$\mathscr{A}$, and its standard deviation,
$\sigma_\mathscr{A}$, follow a monotonous increasing behavior 
with $\sigma_\ell$, as illustrated in 
Figure \ref{fig:anisotropia_macroscopica_sigma_length}(c-d).
Again, the strategy that we use to study the influence of $\sigma_\ell$ on 
the electrical anisotropy of the composite material is to evaluate 
curves of $\wp^{HX}(\Phi)$ for different values of $\sigma_\ell$.
Figure \ref{fig:contour_plot_ell_sigmaell}(b) shows the probability
$\wp^{HX}(\Phi)$ as a contour plot versus $\Phi$ and $\sigma_\ell$. 
It can be seen that the range of values of $\Phi$ for which
$\wp^{HX}(\Phi)\approx 1$ varies very little with the studied 
parameter. 
Only a small increase with increasing $\sigma_\ell$ from 0 mm to 1 mm 
is observed.
Above those values (not shown) practically no variation with $\sigma_\ell$
is observed, and therefore the location and size of the safety zone 
becomes quite insensitive to $\sigma_\ell$.

\subsection{Influence of $\langle\ell\rangle$}

\begin{figure} %11
\centerline{\includegraphics[width=10cm]{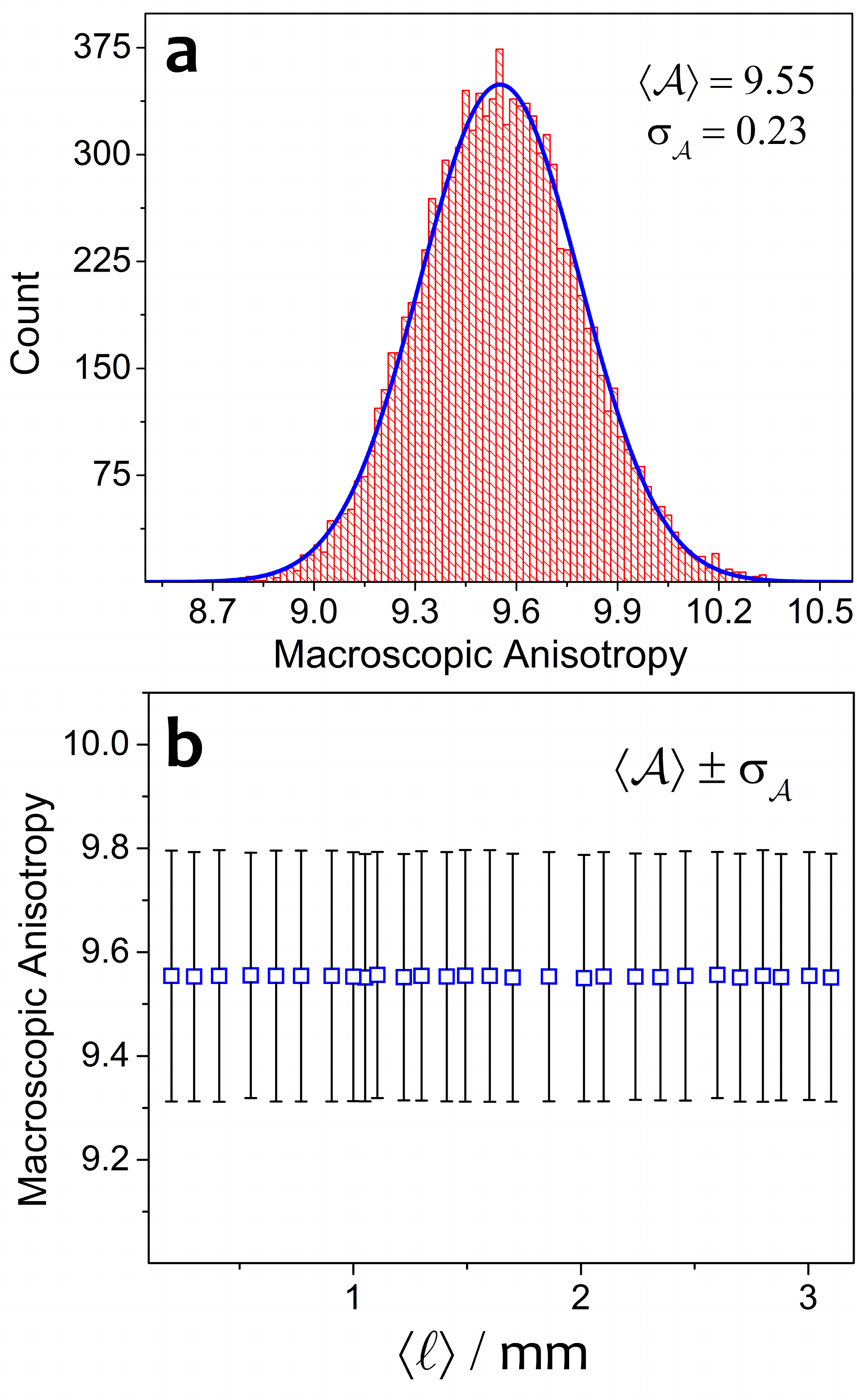}}
\caption{(a) Typical histogram of the macroscopic anisotropy $\mathscr{A}$ 
obtained for a rectangular system of aspect ratio 
$r=L_x/L_y=3/4$, $\mathscr{N}=1000$, and parameters
$\sigma_\theta= 7.5^{\circ}$ and $\sigma_\ell=0.26 \,\text{mm}$.
(b) Mean value of macroscopic anisotropy and its standard deviation 
versus $\langle\ell\rangle$.}
\label{fig:anisotropia_macroscopia_length_media}
\end{figure}

In this case, a Gaussian angular distribution with parameters 
$\langle\theta\rangle=0^\circ$ and $\sigma_\theta=7.5^\circ$, 
and a log-normal distribution for the lengths 
with parameters $\sigma_\ell=0.26\,\text{mm}$ and different 
values of $\langle\ell\rangle$ were assumed.
Figure \ref{fig:contour_plot_ell_sigmaell}(a) shows the probability
$\wp^{HX}(\Phi)$ as a contour 
plot versus $\Phi$ and $\langle\ell\rangle$.
It can be seen that the range of values of $\Phi$ for 
which $\wp^{HX}(\Phi)\approx 1$ varies very little with 
the studied parameter but the value of $\Phi$ required to reach 
the safety zone strongly increases with decreasing 
$\langle\ell\rangle$.
Figure \ref{fig:anisotropia_macroscopia_length_media}(a) shows a typical 
macroscopic anisotropy histogram 
obtained for $\langle\ell\rangle=1.22\,\text{mm}$ by performing 
10500 repetitions, with $\mathscr{N}=1000$. 
For all the values of $\langle\ell\rangle$, the distribution 
is approximately Gaussian, 
with an excellent degree of adjustment, $R^2\geq0.99977$ [continuous-line 
in Figure \ref{fig:anisotropia_macroscopia_length_media}(a)].
Contrary to what was observed for the other two structural 
parameters, in this case the histograms do not change appreciably 
for different values of $\langle\ell\rangle$, as can be clearly seen 
in Figure \ref{fig:anisotropia_macroscopia_length_media}(b), 
in which mean values of macroscopic anisotropy and its standard 
deviation are plotted versus $\langle\ell\rangle$.

\section{Conclusions}

Motivated by experimental work on structured magnetorheologial elastomers,
we present a comprehensive study of stick percolation
in two dimensional networks.
In order to extract realistic parameters for our simulations, we
first carry out a statistical characterization of the distribution 
of metallic sticks in our previously studied MRE samples.
We found that the population of sticks has a log-normal distribution
of stick lengths (centered around $1.35\,\text{mm}$) 
and a Gaussian angular distribution.
The latter is centered around a preferential axis that is given by
the curing magnetic field applied during the sample preparation,
and has a typical standard deviation of approximately five degrees.
In order to simulate the experimental systems, we adopted the model 
of two-dimensional stick percolation and developed a Monte Carlo
numerical algorithm and a computer program implemented in the computer 
language SAGE.
We thoroughly tested our program by reproducing theoretical 
key results of the known scaling behavior of the percolation 
probability in square, isotropically distributed systems.
We then performed extensive numerical simulations of asymmetric 
(rectangular), anisotropic (in the orientation of the sticks) systems, 
modeled after the examined experimental samples.
The main objective of the study was to analyze the effect 
of key structural parameters of the material, which characterize 
the angular and length distribution of the sticks 
(the average length of the sticks $\langle\ell\rangle$, 
the standard deviation of the length distribution $\sigma_\ell$,
and the standard deviation of the angular distribution $\sigma_\theta$) 
on the observation of total electrical anisotropy (TEA).
From a practical point of view, TEA is a
crucial aspect in the design of nano or micro-scale devices like 
pressure mapping sensors and two-dimensional aniso connectors (e.g.
Zebra\textsuperscript{\textregistered}-like connectors for 
parallel flip-chip connections).
We studied the TEA by computing various 
probabilities, especially the only-horizontal probability 
percolation function, $\wp^{HX}$, and analyzing the macroscopic
anisotropy, which quantifies the macrosocpic average degree of orientation 
of the stick population.
We find prescriptions to achieve ``safe'' structural conditions of
total electrical anisotropy, and thus hope to guide the experimentalist 
and technologist to choose the experimental conditions needed 
to make a device with the desired electrical properties. 
Most importantly, we show that there exists a strong dependence of
the TEA on the standard deviation of 
the angular distribution and on the average
length of the sticks, while the standard deviation of the 
length distribution has little effect.

RMN and PIT are research members of the National Council of Research 
and Technology (CONICET, Argentina). 
Financial support was received from UBA (UBACyT projects 2012-2015, 
number 20020110100098 and 2011-2014 number 20020100100741),
and from the Ministry of Science, Technology, and Innovation (MINCYT-FONCYT, 
Argentina, PICT 2011-0377). 
Support from the Center of Documental Production (CePro) and the 
Center of Advanced Microscopy (CMA) (FCEyN, UBA) to obtain the 
shown SEM-TEM images
and from the Low-Temperatures Laboratory (Department of Physics, FCEyN, 
UBA) to implement the device used to generate the structured materials
is gratefully acknowledged. 
We thank the SAGE users community for fruitful discussions 
and Prof.\ William Stein (University of Washington) for 
allowing us to run some of our simulations on SageMathCloud.

%%%%%%%%%%%%%%%%%%%%%%%%%%%%%%%%%%%%%%%%%%%%%%%%%%%%%%%%%%%%%%%%%%%%%%%%%%%%%%%

\end{document}